\documentclass[prc,twocolumn,superscriptaddress]{revtex4}
\usepackage{amsmath,amssymb,epsfig,bm,mathrsfs,feynmp,color}
\usepackage{slashed}
\newcommand{\be}{\begin{equation}}
\newcommand{\ee}{\end{equation}}
\newcommand{\bea}{\begin{eqnarray}}
\newcommand{\eea}{\end{eqnarray}}

\begin{document}
\title{Strange quark matter in strong magnetic fields within a confining model}

\author{Monika Sinha}
\affiliation{Institute for Theoretical
   Physics, J.~W.~Goethe University, D-60438 Frankfurt-Main, Germany}

\author{Xu-Guang Huang}
\affiliation{Center for Exploration of Energy and Matter and
Physics Department, Indiana University, Bloomington, Indiana 47408, USA}

\author{Armen Sedrakian}
\affiliation{Institute for Theoretical
Physics, J.~W.~Goethe University, D-60438 Frankfurt-Main, Germany}

\begin{abstract}                 
  We construct an equation of state of strange quark matter in strong
  magnetic field within a confining model. The confinement is modeled
  by means of the Richardson potential for quark-quark interaction
  modified suitably to account for strong magnetic field. We compare
  our results for the equation of state and magnetization of matter to
  those derived within the MIT bag model. The differences between these
  models arise mainly due to the momentum dependence of the strong
  interaction between quarks in the Richardson model.  Specifically,
  we find that the magnetization of strange quark matter in this model
  has much more pronounced de Haas-van Alf\'{v}en oscillations than in the
  MIT bag model, which is the consequence of the (static)
  gluon-exchange structure of the confining potential.
\end{abstract}

\maketitle

\section{Introduction}\label{intro}

Compact stellar objects can be tentatively divided into two broad
classes: one includes stars made of the ordinary baryonic matter
either in the confined (hadronic) or deconfined (quark-gluon) state,
the second includes stars made of strange matter. The latter
possibility goes back to Witten's idea~\cite{witten} that the
deconfined quark matter composed of an equal number of  up, 
  down and strange quarks may be the true ground state of
matter at high density. Since then, the possibility of strange quark
matter (SQM) and strange stars  made of SQM, as an alternative to
hadronic/quark compact objects, has been continuously explored.

Soft $\gamma$-ray repeaters and anomalous x-ray pulsars are commonly
identified with compact stars with surface magnetic fields $B_s\sim
10^{14}-10^{15}$ G. These objects, which feature the largest
stationary $B$ fields observed in Nature to date, are collectively
termed as ``magnetars.''  The interpretation of astrophysical
manifestations of magnetars requires good knowledge of the properties
of dense matter in the presence of a large magnetic field. There have
been some recent advances in this context in our understanding of the
properties of strange quark matter in strong magnetic fields. The
stationary properties, hydrodynamics, transport, and macroscopic
dynamics have been studied in
Refs.~\cite{huang,HunagKubo,Wen:2012jw,isayev,bordbar,ahmedov}. More
general but related aspects of the physics of fermionic (quark)
matter in strong fields have been discussed recently in, e.g.,
Refs.~\cite{sinha,ferrer,dexheimer,menezes,yue,endrodi}.

In the present work we study the effect of a large magnetic field on
SQM. The properties of cold quark matter at large baryon density is
poorly known due the nonperturbative nature of quantum chromodynamics
(QCD) at densities and temperatures relevant for compact stars. Because
the {\it ab initio} lattice calculations at low temperatures and finite
chemical potentials presently encounter serious problems, effective
phenomenological models are commonly used. Among the the most popular
ones are the MIT bag model~\cite{chodosetal} and the
Nambu-Jona-Lasinio (NJL) model \cite{njl}. Both models have some
merits and some disadvantages. For example, the NJL model exhibits
chiral symmetry breaking, but does not account for the confinement
property of QCD. On the other hand, the bag models are built to
confine through the introduction of an {\it ad hoc} bag pressure but are
unable to account for the chiral symmetry breaking. An alternate to
the bag model way to introduce the confinement is to take 
density-dependent quark masses.  Many phenomenological models have been
proposed in the past which are based on density dependent quark masses
\cite{fowler, chak, d98, lxl}.  We will base our discussion of quark
matter in a strong magnetic field on one such model, that was originally
introduced by Dey {\it et al.}~\cite{d98}.  In this model, the quarks
interact among themselves through the Richardson potential
\cite{rich}, in which the asymptotic freedom and confinement is built
in. Initially, it was used in the meson phenomenology and later tested
in the baryon sector~\cite{ddl}.  This latter model will serve  as a
basis for studying confining strange matter at nonzero temperatures.

Substantial changes in the strange matter properties
  appear when the electromagnetic scales become of the order of the
  nuclear scales, which is the case for fields $B\ge 10^{18}$~G. Such
  fields have not been observed directly in astrophysics, but theoretical
  extrapolations of surface fields observed in magnetars suggest that
  the fields of this magnitude can be reached in the deep interiors of
  compact objects. An upper value of the $B$ field is set by the
  equilibrium that can be sustained by the gravitational forces and
  pressure components of matter in a strong magnetic field. The
  anticipated value of the maximal field is in the range $10^{18}\le
  B_{\rm max }\le 10^{20}$~G, but the precise value of $B_{\rm max}$
  remains uncertain~(Ref.~\cite{ferrer} and references therein).

This work is organized as follows.  In Sec.~\ref{model} we introduce
the Richardson-potential model (hereafter RP model) and demonstrate
its modifications due to the strong magnetic fields. The results of our
numerical computations are shown in Sec.~\ref{results}. Finally, our
findings are summarized in Sec.~\ref{summary}.

\section{Model}\label{model}

We consider SQM in a strong magnetic field at
high densities and nonzero temperature. The $u$, $d$, and $s$ quarks
interact via the Richardson potential~\cite{rich}
\begin{equation}
\label{richpot}
V(q^2) =- \frac49  ~\frac \pi {{\rm ln}[1+(q^2+m_g^2)/\Lambda^2]} \frac{1}{(q^2+m_g^2)},
\end{equation}
where $m_g$ is gluon mass and $\Lambda$ is a scale parameter.
The finite gluon mass is responsible for screening in medium and is
related to the screening length $D$ via
\begin{equation}
 m_g^2 = D^{-2} = \frac{2\alpha_0}{\pi}\sum_{i=u,d,s} k_F^i \mu_i^*,
\end{equation}
where $\alpha_0$ is the perturbative quark gluon coupling,
$\mu_i^* \equiv \sqrt{(k^i_F)^2 + m_i^2}$, $k^i_F$ is the Fermi
momentum, and $m_i$ the quark mass.  The index $i$ labels quark flavors.
An important  feature of our model is that quark masses depend on the density.
We parametrize this dependence as
\begin{equation}
\label{eq:masses}
m_i = M_{i} + M_q\, {\rm sech} \left(\nu \frac{n_b}{n_0}
\right), \qquad i=u,d,s,
\end{equation}
where $n_b = (n_u + n_d + n_s)/3$ is the baryon number density, $n_0$
is the normal nuclear matter density and $\nu$ is a parameter.  At
large $n_b$ the second term in (\ref{eq:masses}) decays exponentially
 and the quark mass $m_i$ falls off from its constituent value
$M_q$ to its current value $M_{i}$.

The number and energy densities of each quark flavor
in the absence of quantizing the magnetic field are given by
\begin{eqnarray}
 n &=& \frac6{(2\pi)^3}\int_0^\infty  f(\epsilon) \, \, \,d^3 k,\\
\varepsilon &=& \frac6{(2\pi)^3}\int_0^\infty f(\epsilon)  \epsilon\, \, \, d^3 k,
\end{eqnarray}
where $\epsilon$ is the single particle energy, $f(\epsilon)=
\{1+\exp[(\epsilon-\mu)/T]\}^{-1}$ is the Fermi distribution function,
with $\mu$ being the chemical potential and $T$ the temperature;
factor 6 is the sum over the spin and color degrees of
freedom. Note that the full single-particle energy $\epsilon$ consists
of the kinetic energy of relativistic particle with mass $m_i$ and the
potential energy arising from the interaction with other quarks via
the Richardson potential (\ref{richpot}).
\begin{figure}[!]
\begin{center}
\vskip 1.cm
\includegraphics[width=8.6cm]{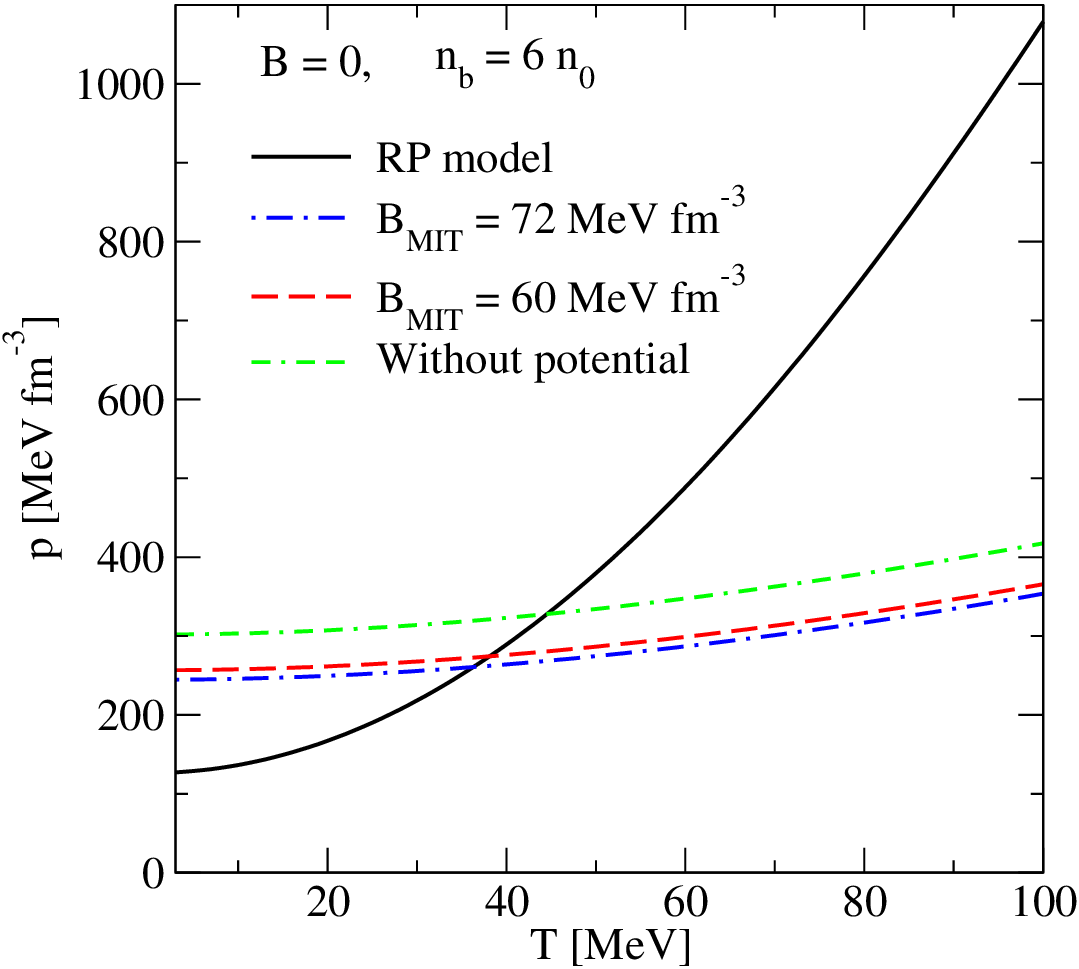}
\vskip 1.2cm
\includegraphics[width=8.6cm]{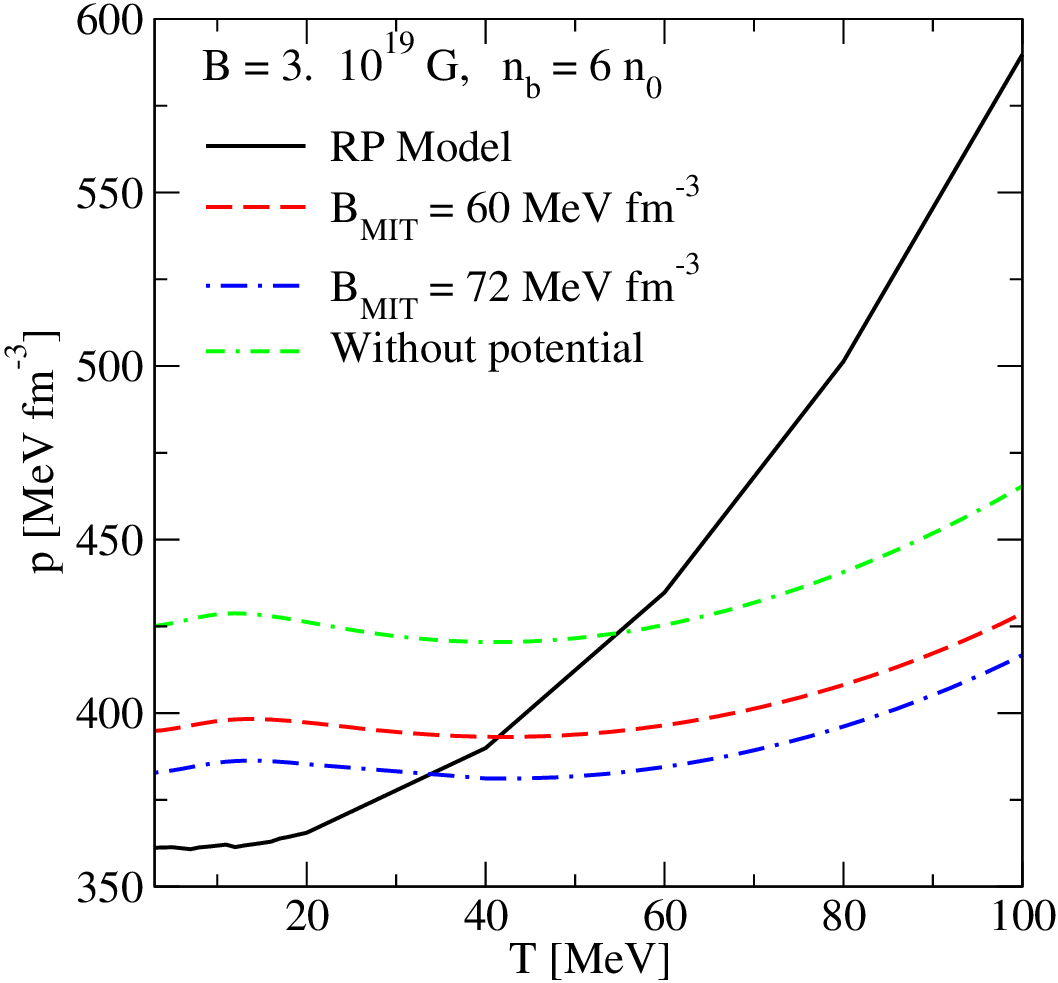}
\caption{(color  online).
Dependence of the thermodynamic pressure $p$ on the temperature at fixed baryon
 number density $n_b=6n_0$ for  the RP model
  (solid, black line), for the MIT bag model with $B_{\rm MIT} = 60$
  MeV fm$^{-3}$ (dashed,  red line) and for $B_{\rm MIT} =
  72$ MeV fm$^{-3}$ (dash-dotted, blue line) and without potential
  (double-dash-dotted, green line). The upper and lower panels correspond to 
the field values  $B=0$  and $B =3\times 10^{19}$~G.
}
\label{temp}
\end{center}
\end{figure}
As is well known, in a magnetic field the motion of charged particles is
Landau quantized in the direction perpendicular to the field. For
sufficiently large magnetic fields one needs to take into account the
modification of the single particle energies and the phase space due
to the Landau quantization of quark orbitals. 

We assume that the field is along the $z$ direction of the Cartesian
coordinate system, ${\mathbf B} = B \hat z$.  Then, the motion is
quantized in the $x$-$y$ plane and the momentum of quarks of mass $m_i$
and charge $eQ_i$ can be decomposed into components parallel and
perpendicular to the $z$ direction, $ {\mathbf k} \equiv (k_z, k_\perp), $
with $ k_\perp^2 = 2ne|Q|B, $ where $e$ is the (positive) unit of
charge.  Consequently, the single particle kinetic energy in the
$n$th Landau level is given by
\begin{equation}
\label{eq:sps}
 \epsilon = \sqrt{k_z^2+m^2+2ne|Q|B}.
\end{equation}
The number density of any  quark flavor is then given by
\begin{equation}
 n = \frac 3{(2\pi)^3} e|Q|B \sum_{n=0}^\infty (2-\delta_{n,0})\int_0^{2\pi}
d\phi \int_{-\infty}^\infty f(\epsilon) \, \, \,dk_z .
\end{equation}
The kinetic part of the energy density for a particular
 quark flavor is given by
\begin{equation}
 \varepsilon_{kin} = \frac 3{(2\pi)^3} e|Q|B
\sum_{n=0}^\infty (2-\delta_{n,0})\int_0^{2\pi}
d\phi \int_{-\infty}^\infty f(\epsilon) \epsilon\, \, \, dk_z.
\end{equation}
The potential part of the energy density due to interaction
between the flavors $i$ and $j$ is given by
\begin{eqnarray}
\varepsilon_{pot}^{ij}
&=& \frac{e^2 |Q_i||Q_j|}{(2\pi)^5}
B^2\sum_{n_i} \sum_{n_j} (2-\delta_{n_i,0}) (2-\delta_{n_j,0})
\nonumber \\
 && \hspace{-1cm}\int_0^{2\pi}\! \! \! d\phi_i \int_0^{2\pi}\!\!   \!  d\phi_j
\int_{-\infty}^\infty  \!  \! \! dk_z^i
 \int_{-\infty}^\infty  \!  \! \! dk_z^j f(\epsilon_i) f(\epsilon_j)NV(q^2)S ,\nonumber\\
\end{eqnarray}
where
\begin{eqnarray}
 N &=& \frac{(\epsilon_i+m_i)(\epsilon_j+m_j)}{4\epsilon_i\epsilon_j},\nonumber\\
S &=& 1+ \frac{k_i^2 k_j^2}{(\epsilon_i+m_i)^2 (\epsilon_j+m_j)^2} 
+ \frac{2{\mathbf k_i}\cdot{\mathbf k_j}}{(\epsilon_i+m_i)(\epsilon_j+m_j)}.\nonumber
\end{eqnarray}
The total energy density is obtained, after summation over the quark flavors,
as
\begin{equation}
\varepsilon = \sum_i\varepsilon_{kin} + \frac12\sum_{i,j} \varepsilon_{pot}^{ij},
\quad \quad
i,j = u, d, s .
\end{equation}
The net entropy density is given by the combinatorial expression for
quark quasiparticles
\bea
s &=& -\frac3{(2\pi)^3} e\sum_i|Q_i|B \sum_{n=0}^\infty \int_0^{2\pi}\!\!
d\phi \int_{-\infty}^\infty\!\! dk_z \nonumber\\
&&\times \{f(\epsilon_i){\rm ln} f(\epsilon_i)
+ [1-f(\epsilon_i)]{\rm ln} [1-f(\epsilon_i)]\},
\eea
where $i$ summation is over the quark flavors.
Then, the thermodynamic pressure is given by
\begin{equation}
\label{eq:pressure}
p = \sum_i \mu_i n_i + Ts -\varepsilon, \quad
i= u, d, s .
\end{equation}
The magnetization of the matter at a given temperature and
constant baryon number density is given by
\begin{equation}
M = \frac{dp}{dB}.
\end{equation}
\begin{figure}[!]
\vskip 1.cm
\begin{center}
\includegraphics[width=8.6cm]{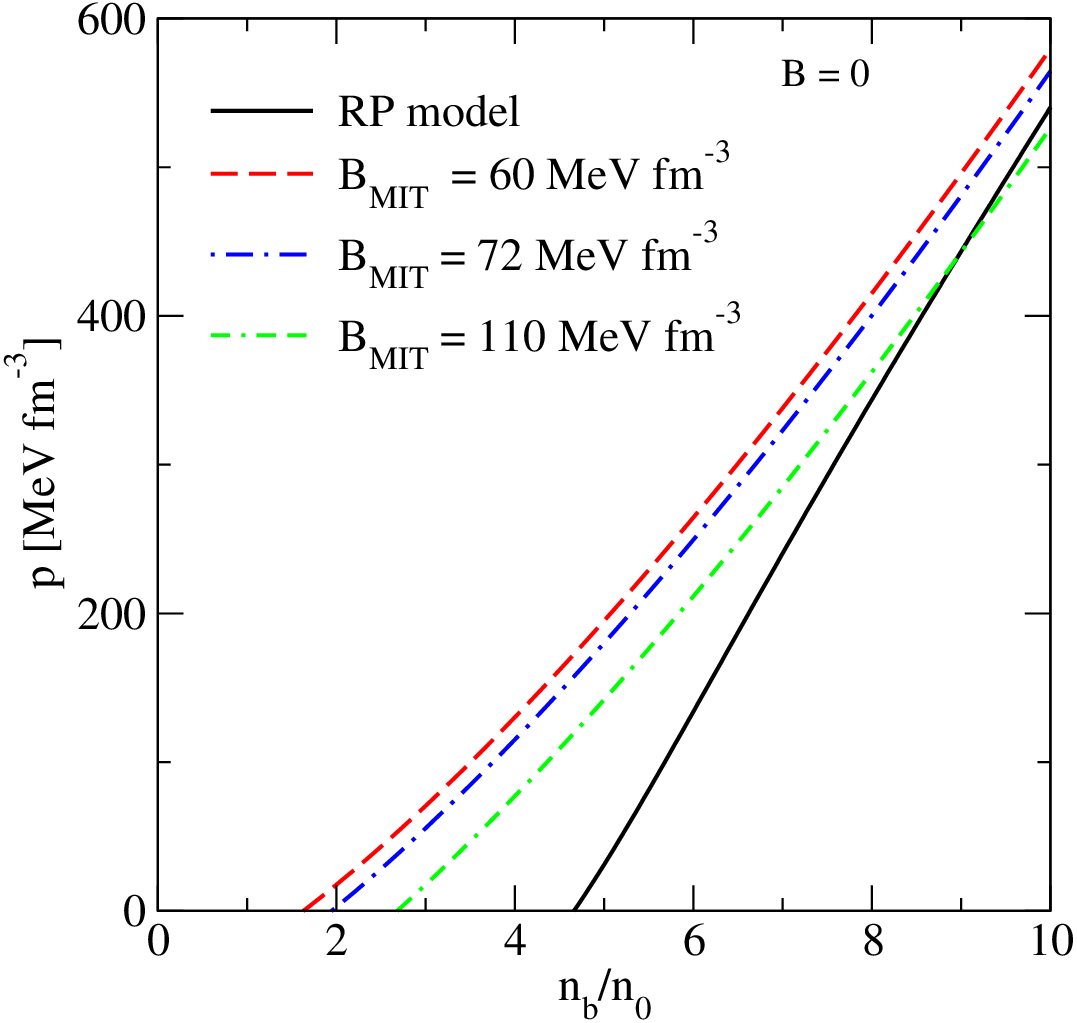}
\vskip 1.2cm
\includegraphics[width=8.6cm]{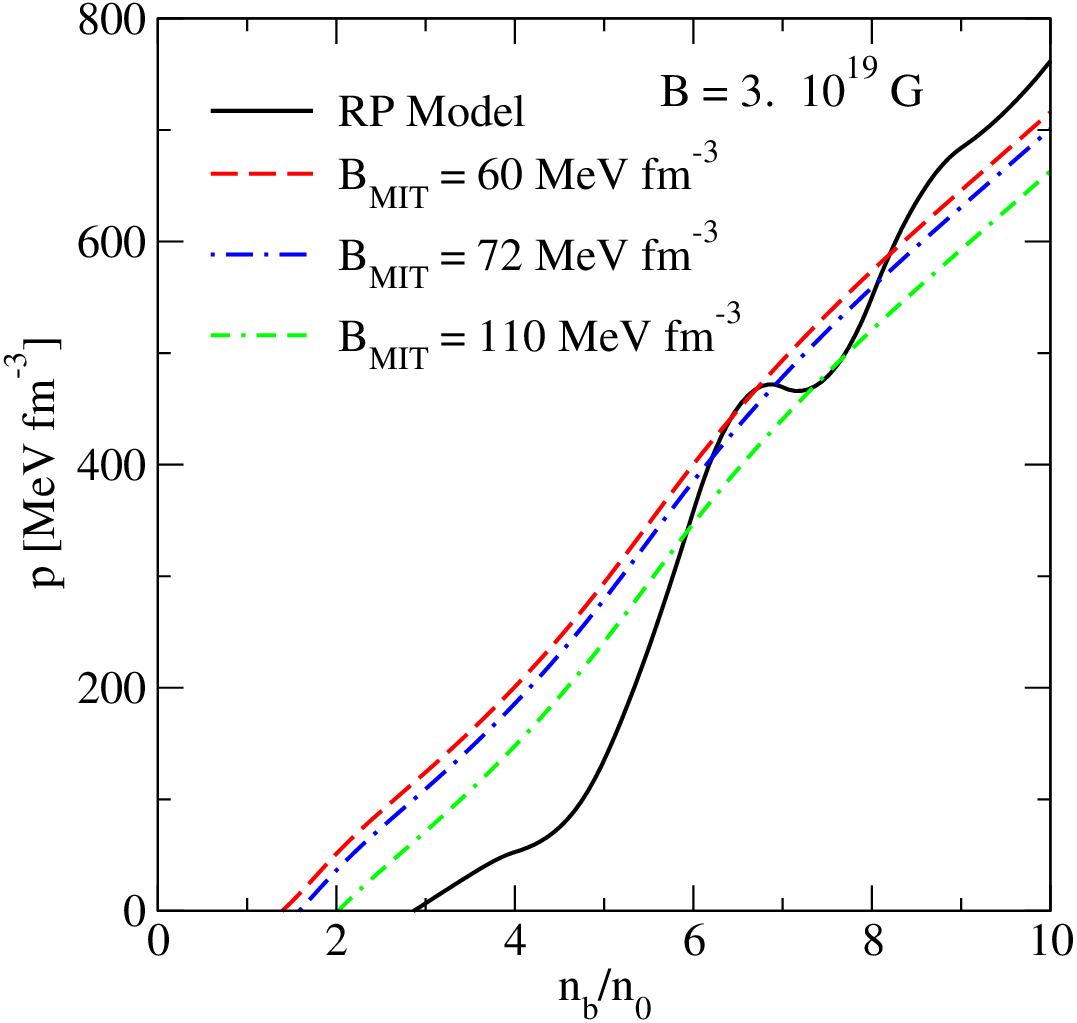}
\caption{(color online).
Dependence  of the thermodynamic pressure $p$ on the normalized baryon
number density  at   $T = 20$ MeV for $B=0$ (upper panel) and $B =3\times
 10^{19}$ G (lower panel).  The pressure is shown for the RP model
 (solid, black line), for the MIT bag model with $B_{\rm MIT} = 60$
MeV fm$^{-3}$ (dashed, red line), $B_{\rm MIT} =
72$ MeV fm$^{-3}$ (dash-dotted, blue line) and $B_{\rm MIT} =
110$ MeV fm$^{-3}$ (double dash-dotted, green line).
}
\label{eos}
\end{center}
\end{figure}
A number of authors~\cite{huang,ferrer,dexheimer,menezes,isayev} have
noticed that  in the presence of a strong magnetic field the pressure is
anisotropic and it is useful to decompose the pressure in components
along ($p_{\parallel}$) and perpendicular ($p_{\perp}$) to the field as
\begin{equation}
p_\parallel = p, \quad  p_\perp= p-MB.
\end{equation}
In strange quark matter the $\beta$ equilibrium can be sustained among
the quark flavors; therefore, the abundances of leptons (electrons and
muons) are negligible. The charge neutrality condition can be written
as
\begin{equation}
\label{eq:charge_neut}
n_u = \frac12(n_d + n_s) = n_b.
\end{equation}
The weak interactions establish an equilibrium among the quark flavors
via the nonleptonic weak process $u + d \rightleftharpoons u +
s \label{reac1}$. Thus, the equilibrium with respect to these weak
reactions requires that the chemical potentials of quark flavors obey
the condition
\begin{equation}
\label{chem1}
\mu_d = \mu_s.
\end{equation}
To summarize, the key equations of our model are Eqs.~(\ref{eq:sps})
-(\ref{eq:pressure}) that are subject to the constraints
(\ref{eq:charge_neut}) and (\ref{chem1}). These equations are solved
self-consistently.
\begin{figure}[t]
\vskip 1cm
\begin{center}
\includegraphics[width=8.6cm]{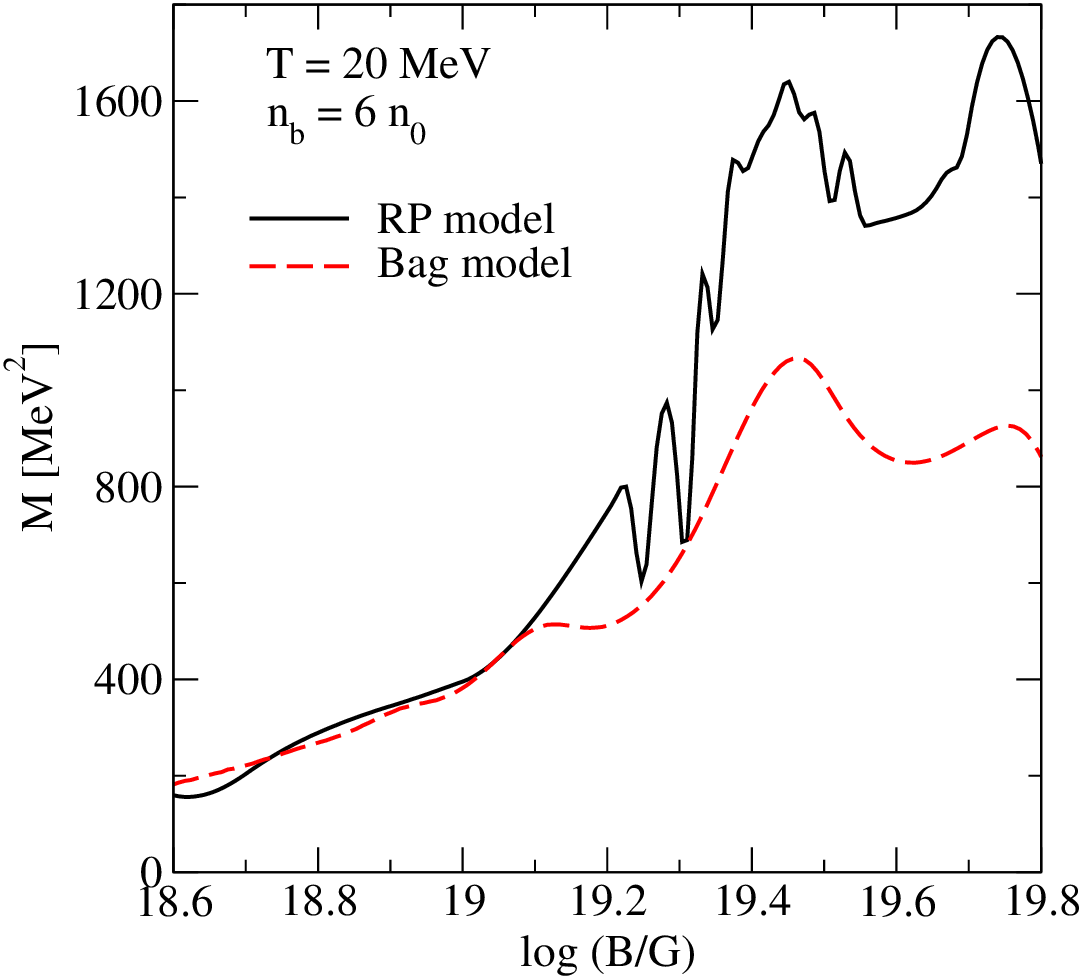}
\caption{(color online).
  Dependence of the magnetization on the magnetic field at
  baryon number density $n_b=6n_0$ and $T=20$ MeV for the RP model
  (solid line, black line) and MIT bag model  (dashed, red line). The
  bag model result does not depend on the value of the bag constant.
}
\label{mlogb}
\end{center}
\end{figure}

\begin{figure}[t]
\vskip 1.cm
\begin{center}
\includegraphics[width=8.6cm]{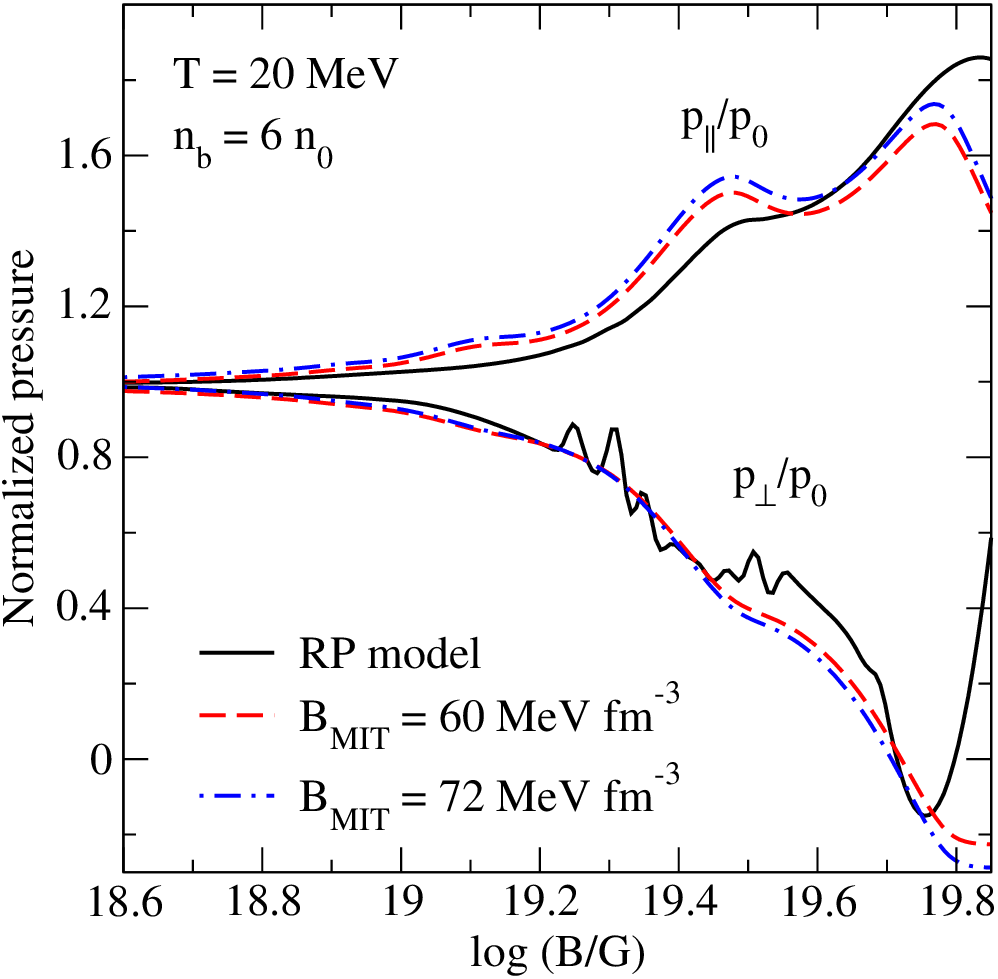}
\caption{
  (color online).
Dependence of the  normalized pressure in parallel and
  perpendicular directions to the magnetic field on the strength of
  the magnetic field at $n_b=6n_0$ and $T=20$ MeV for various models
  (the curves are labeled as in Fig.~\ref{eos}). The upper three
  curves correspond to the parallel pressure and the lower three curves to the
  perpendicular pressure.}
\label{bothp}
\end{center}
\end{figure}

\section{Results}\label{results}

In this section we discuss the results of a numerical solution of the
self-consistent equations presented above. Our main focus will be the
effect of the Richardson potential on the properties of strange matter
in strong magnetic fields at finite temperature. The numerical values
of the parameters of our model are  $\Lambda = 100$ MeV, $\nu =
0.333$, $\alpha_0 = 0.2$, $M_q = 310$, $M_u = 4$, $M_d = 7$, and $M_s = 150$
with all masses given in MeV. A discussion of the feasible parameter
space can be found in Ref.~\cite{d98}.

In Fig.~\ref{temp} we show the function $p(T)$ for the RP model
together with the results obtained with the MIT bag model with two
values of the bag constant along with the result for noninteracting
matter. The case of nonmagnetized and strongly magnetized matter ($B
= 3\times 10^{15}$ G) are displayed.

The bag model and noninteracting gas results are self-similar,
because they differ only by a temperature-independent constant.  In
the absence of a magnetic field the pressure shows $T^2$ power-law behavior
with temperature. In the magnetic field the temperature dependence is
nonmonotonic in the bag model, but in the RP model the temperature
dependence shows the same features as in the absence of a magnetic
field.

In Fig.~\ref{eos} we show the equation of state of SQM in the RP model
and the bag model for fixed $T= 20$ MeV. The bag model equations of
state show $p\propto n^{4/3}$ scaling inherent to the
ultrarelativistic noninteracting gas. In the case of the RP model
the scaling is different because the Richardson potential introduces
additional momentum dependence in the single particle energies, which
results in nearly linear dependence of pressure of density.
Furthermore, in the absence of  a magnetic field the equation of state in
the RP model is softer than in the bag model at low densities and
reaches asymptotically the equation of state with $B_{\rm MIT} = 110$
at high densities.  While the high values of bag constant can mimic
the RP model, for such large values of $B_{\rm MIT}$ the strange
matter is not the absolute ground state of matter.

The upper and lower panels display the differences arising
due to the strong magnetic field ($B = 3\times 10^{19}$ G).  The magnetic
field introduces some oscillations in the pressure with density; in
each case the increase of the pressure after a plateau is caused by
the opening of a new Landau level. For the bag model equation of state
this is more pronounced for the case with the bag value $B_{\rm MIT} =
60$ MeV fm$^{-3}$. The oscillations are much stronger in the RP model
and this can be traced back to the momentum dependence of the
potential. The major contribution comes from the static gluon
propagator part of the potential [the term $(q^2+m_g^2)^{-1}$], while
the logarithmic factor in the potential weakly depends on
momentum. Note that at some density the pressure has a plateau and
slight negative downturn, which can be interpreted as an instability
of homogeneous magnetized matter towards phase separation.

Figure~\ref{mlogb} displays the magnetization of matter as a function of
the magnetic field for the bag model and RP model at fixed $n = 6 n_0$
and $T=20$ MeV. Note that the magnetization does not depend on the bag
constant. For fields $B> 10^{19}$ G the magnetization shows de Haas -
van Alf\'{v}en oscillations in both models. However, the oscillations are
much more pronounced in the RP model than in the bag model. This is
the consequence of the momentum dependence of the RP interaction,
which has the structure of the static gluon exchange. A similar effect
was observed in Ref.~\cite{huang} in a noninteracting strange quark
matter model. Note also, the absolute value of the magnetization is by
a factor 2 lager in the RP model for sufficiently large fields.

At large magnetic fields the anisotropy due to the magnetic field is
important. The pressure components in parallel and perpendicular
direction to the magnetic field are not the same. We show the
variations of $p_\parallel$ and $p_\perp$ with $B$ in the RP and bag
models at $n=6n_0$ in Fig.~\ref{bothp}. We note that below
$B=3\times10^{18}$ G, both $p_\parallel$ and $p_\perp$ are practically
equal to the pressure of matter in absence of a magnetic field. Hence,
for the SQM with the model under consideration the effect of the magnetic
field is not significant below $B\sim 10^{18}$ G. With the increase of
$B$, $p_\parallel$ increases whereas $p_\perp$ decreases for both
models. For large fields, at a certain value of $B$, $p_\perp$ becomes
negative and this critical value is almost the same in both
models. Recalling that without the confining potential, at a very large
magnetic field $p_\perp\rightarrow 0$~\cite{huang,dexheimer}, we see
that the confining potential provides additional ``attraction" inside
the SQM, and its effect becomes more transparent at larger $B$.  The
oscillations of the function $p_\perp$ reflect the oscillations in the
magnetization.
\begin{figure}[t]
\vskip 1.cm
\begin{center}
\includegraphics[width=8.6cm]{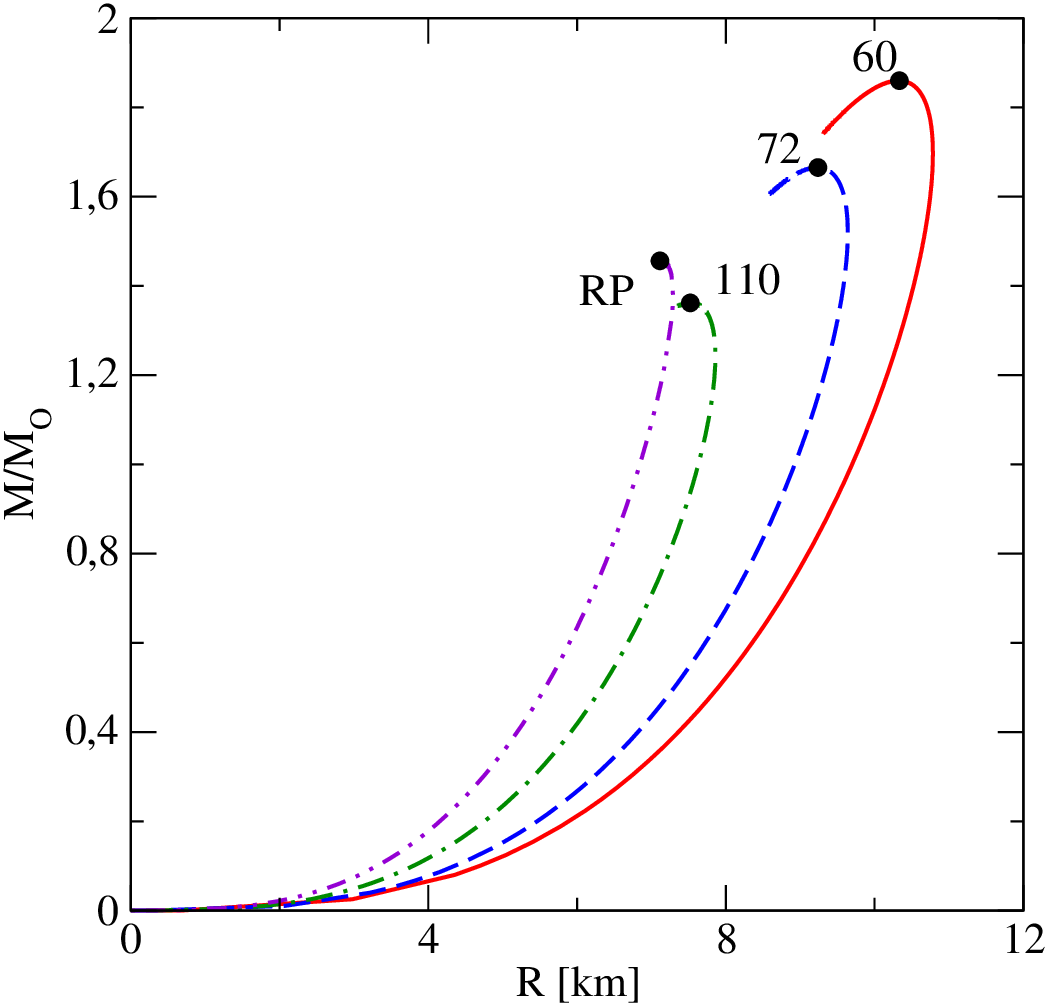}
\caption{(color online). Mass-radius relation for strange stars in the
  absence of magnetic fields.  The MIT bag model based results are
  labeled by the value of the $B_{\rm MIT}$; those based on the RP model by
  ``RP''.    }
\label{MR_relation}
\end{center}
\end{figure}
The mass-radius relation for the underlying models in the absence of
a magnetic field are shown in Fig.~\ref{MR_relation}, which demonstrates
the key difference between the RP and bag models in the astrophysics
context. Because of the softer equation of state of the RP model the
strange stars are more compact (the radii are smaller) and their
maximum mass is by about $20\%$ smaller than for the models with
$B_{\rm MIT}\sim 60-70$ MeV fm$^{-3}.$ The computation of the
mass-radius relation in the case of strongly magnetized matter can be
carried out on the basis of the equations of state obtained in this
work. Such calculation requires the solution of Einstein's equations
in axial symmetry, because of the anisotropy in the pressure induced
by the magnetic field and is beyond the scope of this work (see
e.g.~\cite{Cardall:2000bs}).

\section{Summary}\label{summary}

In this work we studied the effects of strong magnetic fields,
quark-quark confining interaction, and chiral symmetry restoration on
the equation of state of the charge neutral strange quark matter. The
confining interaction is modeled by the Richardson potential (RP)
which features both the asymptotic freedom and the confinement. The
chiral symmetry restoration is parametrized as a smooth crossover of
the quark masses from their constituent values at low baryon densities
to their current ones at large baryon densities. We compared the RP
model to the MIT bag model. We find significant differences between
the equation of state and the magnetization of the strange quark
matter predicted by these models. This is the result of the intrinsic
momentum dependence in the interaction of the RP model, which mimics 
the one-gluon-exchange interaction of the
QCD. Specifically, we find that (a) the thermodynamic pressure in the RP
model is more sensitive to temperature and baryon density when the
magnetic field is strong; (b) the magnetization is larger in the RP
model than in the bag model in the limit of large fields, $B>10^{19}$
G; (c) the de Haas-van Alf\'{v}en oscillations in the magnetization and in
the transverse pressure $p_\perp$ is more pronounced in the RP model.

Furthermore, we find that the presence of a confining potential,
modeled either in terms of the RP potential or the MIT bag, suppresses
the pressure components $p_\parallel$ and $p_\perp$ and, at large $B$,
the anisotropy in the equation of state. The splitting between the
longitudinal pressure $p_\parallel$ and the transverse pressure
$p_\perp$ was found to be weaker than that in free (noninteracting)
SQM. This underlines the importance of taking into account the
confining potential in studies of strongly magnetic SQM matter in
cores of neutron stars and in strange stars. It remains an interesting
task to explore the effects of the confining potential in a  strong
magnetic field on the structure and geometry of such stars.

 The strong magnetic fields in the interiors of
  strange stars will affect the transport process and weak interaction
  rates. The strong de-Haas--van Alf\'{v}en oscillations in the magnetic
  field will induces oscillations in, for example, the transport
  coefficients, as demonstrated for the bulk viscosity in
  Ref.~\cite{huang}. They will affect the kinematics of Urca
  processes, as in the case of nucleonic matter~\cite{Baiko:1998jq}
  and may open an additional channel of neutrino bremsstrahlung due
  the Pauli paramagnetic shift in the Fermi levels of
  quarks~\cite{vanDalen:2000zw}.

  In this work we assumed that the strange matter is in the normal
  (unpaired) state. It is likely that the flavor symmetric quark
  matter at low temperatures will be a superfluid. The interplay
  between the superfluidity and magnetism in quark matter has been
  studied in a number of contexts~\cite{alford,PerezMartinez:2011zzb,
    Wu:2011qj,Yu:2012jn,Feng:2012dqa,Mandal:2012fq,Noronha:2007wg};
  however, much remains still unexplored, one possible subject being
  the extension of the present setup to the case of superfluidity of
  strange matter.

\section*{Acknowledgements} We thank D. H. Rischke for discussions.
M. S. acknowledges the support of the Alexander von Humboldt
Foundation. X.-G. H. acknowledges the support from Indiana University
Grant 22-308-47 and the US DOE Grant DE-FG02-87ER40365.


\begin{thebibliography}{99}
\bibitem{witten}
  E.~Witten,
  Phys.\ Rev.\ D {\bf 30}, 272 (1984).

\bibitem{huang}
  X.~-G.~Huang, M.~Huang, D.~H.~Rischke, and A.~Sedrakian,
  Phys.\ Rev.\ D {\bf 81}, 045015 (2010).

\bibitem{HunagKubo}  X.~-G.~Huang, A.~Sedrakian, and D.~H.~Rischke,
Ann. Phys. (NY)\  {\bf 326}, 3075 (2011).

\bibitem{Wen:2012jw}
  X.~-J.~Wen, S.~-Z.~Su, D.~-H.~Yang and G.~-X.~Peng,
  Phys.\ Rev.\ D {\bf 86} (2012) 034006
  [arXiv:1207.6148 [hep-ph]].

\bibitem{isayev}
 A.~A.~Isayev and J.~Yang,
 J.\ Phys.\ G {\bf 40}, 035105 (2013).

\bibitem{bordbar}
 G.~H.~Bordbar, F.~Kayanikhoo, and H.~Bahri,
Iranian J. Sci. Tech. A {\bf 37}, 165 (2013).


\bibitem{ahmedov}
B.~J.~Ahmedov, B.~B.~Ahmedov, and A.~A.~Abdujabbarov,
Astrophys.\ Space Sci.\  {\bf 338}, 18 (2012).

\bibitem{sinha}
  M.~Sinha and D.~Bandyopadhyay,
  Phys.\ Rev.\ D {\bf 79}, 123001 (2009);
  R.~Mallick and M.~Sinha,
  Mon.\ Not.\ R.\ Astron.\ Soc.\  {\bf 414}, 2702 (2011);
  M.~Sinha, B.~Mukhopadhyay, and A.~Sedrakian,
  Nucl.\ Phys.\ A {\bf 898}, 43 (2013);
  B.~Mukhopadhyay and M.~Sinha,
  arXiv:1302.3444.

\bibitem{ferrer}
  E.~J.~Ferrer, V.~de la Incera, J.~P.~Keith, I.~Portillo, and P.~L.~Springsteen,
  Phys.\ Rev.\ C {\bf 82}, 065802 (2010);
  L.~Paulucci, E.~J.~Ferrer, V.~de la Incera, and J.~E.~Horvath,
  Phys.\ Rev.\ D {\bf 83}, 043009 (2011);
  E.~J.~Ferrer and V.~de la Incera,
Lect. Notes Phys. {\bf 871}, 399 (2013).


\bibitem{dexheimer}
  V.~Dexheimer, R.~Negreiros, and S.~Schramm,
  Eur.\ Phys.\ J.\ A {\bf 48}, 189 (2012);
  M.~Strickland, V.~Dexheimer, and D.~P.~Menezes,
  Phys.\ Rev.\ D {\bf 86}, 125032 (2012);
  V.~Dexheimer, R.~Negreiros, S.~Schramm, and M.~Hempel,
  arXiv:1208.1320;
  V.~Dexheimer, D.~P.~Menezes, and M.~Strickland,
  arXiv:1210.4526.

\bibitem{menezes}
  D.~P.~Menezes, M. B. Pinto, S.~S.~Avancini, A.~P. Martinez, and C.~Providencia,
  Phys.\ Rev.\ C {\bf 79}, 035807 (2009);
  S.~S.~Avancini, D.~P.~Menezes, and C.~Providencia,
  Phys.\ Rev.\ C {\bf 83}, 065805 (2011);
  A.~Rabhi, P.~K.~Panda, and C.~Providencia,
  Phys.\ Rev.\ C {\bf 84}, 035803 (2011);
  A.~Rabhi and C.~Providencia,
  Phys.\ Rev.\ C {\bf 83}, 055801 (2011).

\bibitem{yue}
  P.~Yue, F.~Yang, and H.~Shen,
  Phys.\ Rev.\ C {\bf 79}, 025803 (2009).

\bibitem{endrodi}
  G.~Endrodi,
J. High Energy Phys. {\bf 04},  023 (2013).


\bibitem{chodosetal}
  A.~Chodos, R.~L.~Jaffe, K.~Johnson, C.~B.~Thorn, and V.~F.~Weisskopf,
  Phys.\ Rev.\ D {\bf 9}, 3471 (1974).

\bibitem{njl}
  Y.~Nambu and  G.~Jona-Lasinio,
  Phys.\ Rev.\  {\bf 122}, 345 (1961); Phys.\ Rev.\  {\bf 124}, 246 (1961).

\bibitem{fowler}
  G.~N.~Fowler, S.~Raha, and R.~M.~Weiner,
  Z.\ Phys.\ C {\bf 9}, 271 (1981).

\bibitem{chak}
  S.~Chakrabarty,
  Phys.\ Rev.\ D {\bf 43}, 627 (1991).

\bibitem{d98}
  M.~Dey, {\it et al.}
  Phys.\ Lett.\ B {\bf 438}, 123 (1998); {\bf 447}, 352 (1999)].

\bibitem{lxl}
  A. Li, R. X. Xu, and J. F. Lu,
  Mon. Not. R. Astron. Soc. {\bf 402}, 2715L (2010).

\bibitem{rich}
  J. L. Richardson,
  Phys. Lett. B {\bf  82}, 272 (1979).

\bibitem{ddl} J. Dey, M. Dey, and J. LeTourneux,
  Phys.\ Rev.\ D {\bf 34}, 2104 (1986).


\bibitem{Cardall:2000bs} 
  C.~Y.~Cardall, M.~Prakash, and J.~M.~Lattimer,
  Astrophys.\ J.\  {\bf 554}, 322 (2001).

\bibitem{Baiko:1998jq} 
  D.~A.~Baiko and D.~G.~Yakovlev,
  Astron.\ Astrophys.\  {\bf 342}, 192 (1999).

\bibitem{vanDalen:2000zw} 
  E.~N.~E.~van Dalen, A.~E.~L.~Dieperink, A.~Sedrakian and R.~G.~E.~Timmermans,
Astron. and Astrophys. {\bf 360}, 549 (2000).




\bibitem{alford}
 M.~G.~Alford and A.~Sedrakian,
 J.\ Phys.\ G {\bf 37}, 075202 (2010).


\bibitem{PerezMartinez:2011zzb}
  A.~Perez Martinez, R.~Gonzalez Felipe and D.~Manreza Paret,
  Int.\ J.\ Mod.\ Phys.\ E {\bf 20}  84  (2011).

\bibitem{Wu:2011qj}
  P.~-p.~Wu, H.~He, D.~Hou and H.~-c.~Ren,
  Phys.\ Rev.\ D {\bf 84}, 027701 (2011) 

\bibitem{Yu:2012jn}
  L.~Yu and I.~A.~Shovkovy,
  Phys.\ Rev.\ D {\bf 85}, 085022 (2012).

\bibitem{Feng:2012dqa}
  B.~Feng, E.~J.~Ferrer and V.~de la Incera,
  Phys.\ Rev.\ D {\bf 85}, 103529 (2012).


\bibitem{Mandal:2012fq} 
  T.~Mandal and P.~Jaikumar,
  Phys.\ Rev.\ C {\bf 87}, 045208 (2013).


\bibitem{Noronha:2007wg} 
  J.~L.~Noronha and I.~A.~Shovkovy,
  Phys.\ Rev.\ D {\bf 76}, 105030 (2007)
  [Erratum-ibid.\ D {\bf 86}, 049901 (2012)]
  [arXiv:0708.0307 [hep-ph]].

\end{thebibliography}
\end{document}